\documentclass[preprint,showpacs,preprintnumbers,amsmath,amssymb]{revtex4}
\usepackage{graphicx}
\usepackage{dcolumn}
\usepackage{bm}

\newcommand{\be}{\begin{equation}}
\newcommand{\en}{\end{equation}}
\newcommand{\bea}{\begin{eqnarray}}
\newcommand{\ena}{\end{eqnarray}}

\begin{document}


\title{Chaplygin inflation on the brane  }

\author{Ram\'on Herrera}
\email{ramon.herrera@ucv.cl} \affiliation{ Instituto de
F\'{\i}sica, Pontificia Universidad Cat\'{o}lica de
Valpara\'{\i}so, Casilla 4059, Valpara\'{\i}so, Chile.}

\date{\today}

\begin{abstract}
 Brane inflationary universe model in the context of a  Chaplygin
 gas equation of state is studied. General
 conditions  for this model to be realizable are
 discussed. In the high-energy limit and by using a chaotic potential we
 describe in great details the characteristic of this model.
  The parameters of the model are restricted by
 using recent astronomical  observations.
\end{abstract}

\maketitle

\section{Introduction}

It is well known that inflation is to date the most compelling
solution to many long-standing problems of the Big Bang model
(horizon, flatness, monopoles, etc.) \cite{guth,infla}. One of the
success of the inflationary universe model is that it provides a
causal interpretation of the origin of the observed anisotropy of
the cosmic microwave background (CMB) radiation, and also the
distribution of large scale structures \cite{astro}.

In concern to higher dimensional theories, implications of
string/M-theory to Friedmann-Robertson-Walker (FRW) cosmological
models have recently attracted  great deal of attention, in
particular, those related to brane-antibrane configurations such
as space-like branes\cite{sen1}. The realization  that we may live
on a so-called brane embedded in a higher-dimensional Universe has
significant implications to cosmology \cite{1}. In this scenario
the  standard model of particle is confined to the  brane, while
gravitations propagate in the bulk spacetimes. Since, the effect
of the extra dimension induces additional terms in the Friedmann
equation  \cite{22,3}. One of the term that appears in this
equation is  a  quadratic term in the energy  density. Such a term
generally makes it easier to obtain inflation in the early
Universe \cite{4,5}. For a review, see, e.g., Ref.\cite{M}.

On the other hand, the generalized Chaplygin gas    has been
proposed as an  alternative model for  describing the accelerating
of the universe. The generalized Chaplygin gas is described by an
exotic equation of state of the form \cite{Bento}
\begin{equation}
 p_{ch} = - \frac{A}{\rho_{ch}^\beta},\label{1}
\end{equation}
where $\rho_{ch}$ and $p_{ch}$ are the energy density and pressure
of the generalized Chaplygin gas, respectively. $\beta$ is a
constant that lies in  the range $0 <\beta\leq 1$, and $A$ is a
positive constant. The original  Chaplygin gas corresponds to the
case $\beta = 1$ \cite{2}. Inserting this equation of state into
the relativistic energy conservation equation leads to an energy
density given by
\begin{equation}
 \rho_{ch}=\left[A+\frac{B}{a^{3(1+\beta)}}\right]^{\frac{1}{1+\beta}},
 \label{2}
\end{equation}
 where $a$ is the
scale factor  and $B$ is a positive integration
constant\cite{Bento}.

The Chaplygin gas emerges as a effective fluid of a generalized
d-brane in a (d+1, 1) space time, where the   action can be
written as a generalized Born-Infeld action \cite{Bento}. These
models have been extensively studied in the literature
\cite{other}. The  parameters of the model were constrained using
currents cosmological observations, such as, CMB \cite{CMB} and
supernova of type Ia (SNIa) \cite{SIa}.

In the  model of  Chaplygin inspired inflation usually the scalar
field, which drives inflation, is the standard inflaton field,
where the energy density given by Eq.(\ref{2}), can be extrapolate
for  obtaining a successful inflation period with a Chaplygin gas
model\cite{Ic}. Recently, tachyon-Chaplygin inflationary universe
model was considered in \cite{yo}, and the dynamics of the early
universe and the initial conditions for inflation in a model with
radiation and a Chaplygin gas was studied  in
Ref.\cite{Monerat:2007ud}.  As far as we know, a Chaplygin
inspired inflationary model in which a brane-world model is
considered has not been studied.

 The motivation for introducing Chaplygin-brane scenarios is
the increasing interest in higher-dimensional cosmological models,
motivated by superstring theory, where the matter fields are
confined to a lower-dimensional  brane(related to open string
modes), while gravity can propagate in the bulk (closed string
modes). On the other hand, the  Chaplygin gas model seems to be a
viable alternative to models that provide    an accelerated
expansion of the early universe.
 Our aim is to quantify the modifications of the Chaplygin
inspired inflation in the brane scenario. In order to do this we
study the early universe dynamic and the cosmological
perturbations in our model. We will show that these underlying
assumptions allow for an inflationary scenario in which the
observational constrains are successfully met.

The outline of the paper is a follows. The next section presents a
short review of the modified Friedmann equation by using a
Chaplygin gas, and  we present the brane-Chaplygin inflationary
model. Section \ref{sectpert} deals with the calculations of
cosmological perturbations in general term.  In Section
\ref{exemple} we use a chaotic potential in the high-energy limit
for obtaining explicit expression for the model. Finally,
Sect.\ref{conclu} summarizes our findings.

\section{The modified Friedmann equation and the brane-Chaplygin Inflationary phase. }

 We consider the five-dimensional brane scenario, in which the
Friedmann equation is modified from its usual form, in the
following way\cite{2,3}
\begin{equation}
H^2=\kappa\,\rho_\phi\left[1+\frac{\rho_\phi}{2\lambda}\right]+\frac{\Lambda_4}{3}+\frac{\xi}{a^4},
\end{equation}
where  $H=\dot{a}/a$ denotes the Hubble parameter, $\rho_\phi$
represents the matter confined to the brane, $\kappa=8\pi
G/3=8\pi/3m_p^2$, $\Lambda_4$ is the four-dimensional cosmological
constant and  the final term represents the influence of bulk
gravitons on the brane, where $\xi$ is an integration constant
(this term appears as a form of dark radiation). The brane tension
$\lambda$ relates the four and five-dimensional Planck masses via
$m_p=\sqrt{3M_5^6/(4\pi\lambda)}$, and is constrained by the
requirement of successful nucleosynthesis as $\lambda >$
(1MeV)$^4$ \cite{Cline}. We assume that the four-dimensional
cosmological constant is set to zero, and once inflation begins
the final term will rapidly become unimportant, leaving us
with\cite{M}

\begin{equation}
H^2=\kappa\left[A+\rho_\phi^{(1+\beta)}\right]^{\frac{1}{1+\beta}}\,
\left[1+\frac{\left[A+\rho_\phi^{(1+\beta)}\right]^{\frac{1}{1+\beta}}}{2\lambda}\right].
\label{HC}
\end{equation}
Here, $\rho_\phi$  becomes
$\rho_\phi=\frac{\dot{\phi}^2}{2}+V(\phi)$,
  and $V(\phi)=V$ is
the scalar  potential. Note that, in the low energy regime
$[A+\rho_\phi^{(1+\beta)}]^{1/(1+\beta)}\ll\lambda$, the standard
Chaplygin inflationary model is recovered, and in a very
hight-energy regime, the contribution from the matter in
Eq.(\ref{HC})  becomes proportional to
$[A+\rho_\phi^{(1+\beta)}]^{2/(1+\beta)}$ in the effective  energy
density.

 We assume that the scalar field is confined
to the brane, so that its field equation has the standard form
 \be \ddot{\phi}\,+3H \;
\dot{\phi}+V'=0, \label{key_01}
 \en
where dots mean derivatives with respect to the cosmological time
and
 $V'=\partial V(\phi)/\partial\phi$. For convenience we will use
 units in which $c=\hbar=1$.

  The modification of the Eq.(\ref{HC}) is realized from an
extrapolation of Eq.(\ref{2}), where  the density matter
$\rho_m\sim a^{-3}$ in introduced in such  a way that we may write
\begin{equation}
 \rho_{ch}=\left[A+\rho_m^{(1+\beta)}\right]^{\frac{1}{1+\beta}}\longrightarrow
 \left[A+\rho_\phi^{(1+\beta)}\right]^{\frac{1}{1+\beta}},\label{extr}
\end{equation}
and  thus, we identifying $\rho_m$ with the contributions of the
scalar field which gives Eq.(\ref{HC}). The generalized Chaplygin
gas model may be viewed as a modification of gravity, as described
in Ref.\cite{Ber}, and for chaotic inflation, in Ref.\cite{Ic}.
Different modifications of gravity have been proposed in the last
few years, and there has been a lot of interest in the
construction of early universe scenarios in higher-dimensional
models motivated by string/M-theory \cite{Ran}. It is  well-known
that these modifications can lead to important changes in the
early universe. In the following  we will take $\beta=1$ for
simplicity, which means the usual Chaplygin gas.

During the inflationary epoch the energy density associated to the
 scalar field is of the order of the potential, i.e.
$\rho_\phi\sim V$. Assuming the set of slow-roll conditions, i.e.
$\dot{\phi}^2 \ll V(\phi)$ and $\ddot{\phi}\ll 3H\dot{\phi}$, the
Friedmann equation (\ref{HC}) reduces  to
\begin{eqnarray}
H^2\approx
\kappa\,\sqrt{A+V^2}\left[1+\frac{\sqrt{A+V^2}}{2\lambda}\right],\label{inf2}
\end{eqnarray}
and  Eq. (\ref{key_01}) becomes
\begin{equation}
3H \dot{\phi}\approx-V'. \label{inf3}
\end{equation}

Introducing the dimensionless slow-roll parameters \cite{4}, we
write
\begin{equation}
\varepsilon=-\frac{\dot{H}}{H^2}\simeq\frac{m_p^2}{16\pi}\,\left[\frac{V\,V'^2}{(A+V^2)^{3/2}}
\,\frac{\left(1+\frac{(A+V^2)^{1/2}}{\lambda}\right)}{\left(1+\frac{(A+V^2)^{1/2}}{2\lambda}\right)^2}\right],\label{ep}
\end{equation}
and
\begin{equation}
\eta=-\frac{\ddot{\phi}}{H\,
\dot{\phi}}\simeq\,\frac{m_p^2}{8\pi}\,\frac{V''}{(A+V^2)^{1/2}}\,\left[1+\frac{(A+V^2)^{1/2}}{2\lambda}
\right]^{-1}\label{eta}.
\end{equation}

Note that in the limit $A\rightarrow 0$, the slow-parameters
$\varepsilon$ and $\eta$ coincides with brane-inflation \cite{4}.
Also, in the low-energy limit, $\sqrt{A+\rho_\phi^2}\ll\lambda$,
the slow-parameters reduce to the standard form \cite{Ic}.

The condition under  which inflation  takes place can be
summarized with the parameter $\varepsilon$ satisfying  the
inequality $\varepsilon<1$, which  is analogue to the requirement
that  $\ddot{a}> 0$. This condition could be written in terms of
the  scalar potential and its  derivative $V'$, which becomes
\begin{equation}
V\,V'^2\,\left[1+\frac{(A+V^2)^{1/2}}{\lambda}\right]<\frac{16\pi}{m_p^2}\,(A+V^2)^{3/2}\,
\left[1+\frac{(A+V^2)^{1/2}}{2\lambda}\right]^2.\label{cond}
\end{equation}

Inflation ends when the universe heats up at a time when
$\varepsilon\simeq 1$, which implies
\begin{equation}
V_f\,V_f'^2\,\left[1+\frac{(A+V_f^2)^{1/2}}{\lambda}\right]\simeq\frac{16\pi}{m_p^2}\,(A+V_f^2)^{3/2}\,
\left[1+\frac{(A+V_f^2)^{1/2}}{2\lambda}\right]^2.\label{al}
\end{equation}
However, in the high-energy limit
$[A+\rho_\phi^2]^{1/2}\approx[A+V^2]^{1/2}\gg\lambda$
Eq.(\ref{al}) becomes
$$
V_f'^2\simeq\frac{4\pi}{m_p^2\,\lambda}\,\frac{(A+V_f^2)^2}{V_f}.
$$

The number of e-folds at the end of inflation is given by
\begin{equation}
N=-\frac{8\pi}{m_p^2}\,\int_{\phi_{*}}^{\phi_f}\frac{\sqrt{A+V^2}}{V'}
\,\left[1+\frac{\sqrt{A+V^2}}{2\lambda}\right]\,d\phi,\label{N}
\end{equation}
or equivalently
\begin{equation}
N=-\frac{8\pi}{m_p^2}\,\int_{V_{*}}^{V_f}\frac{\sqrt{A+V^2}}{V'^2}\left[1+\frac{\sqrt{A+V^2}}{2\lambda}\right]
\,d\,V.\label{NV}
\end{equation}
Note that in the high-energy limit Eq.(\ref{NV}) becomes
$N\simeq-(4\pi/m_p^2\lambda)\int_{V_*}^{V_f}\,[(A+V^2)/V'^2]dV$.

  In the following, the subscripts  $*$ and $f$ are
used to denote  the epoch when the cosmological scales exit the
horizon and the end of  inflation, respectively.

\section{Perturbations\label{sectpert}}

In this section we will study the scalar and tensor perturbations
for our model. It was shown in Ref. \cite{PB} that the
conservation of the curvature perturbation, $\cal{R}$, holds for
adiabatic perturbations irrespective of the form of gravitational
equations by considering the local conservation of the
energy-momentum tensor. One has
${\cal{R}}=\psi+H\delta\phi/\dot{\phi}\simeq
(H/\dot{\phi})(H/2\pi)$, where $\delta\phi$ is the perturbation of
the scalar field $\phi$. For a scalar field the power spectrum of
the curvature perturbations  is given  in the slow-roll
approximation by following expression \cite{4}

\begin{equation}
{\cal{P}_R}\simeq\left(\frac{H^2}{2\pi\dot{\phi}}\right)^2\,\simeq\,\frac{128\pi}{3m_p^6}\,\frac{(A+V^2)^{3/2}}{V'^2}\,
\left[1+\frac{(A+V^2)^{1/2}}{2\lambda}\right]^3 .\label{dp}
\end{equation}
Note that in the limit $A\rightarrow 0$ the amplitude of scalar
perturbation given by Eq.(\ref{dp}) coincides with Ref.\cite{4}.

The scalar spectral index $n_s$ is given by $ n_s -1 =\frac{d
\ln\,{\cal{P}_R}}{d \ln k}$,  where the interval in wave number is
related to the number of e-folds by the relation $d \ln k(\phi)=-d
N(\phi)$. From Eq.(\ref{dp}), we get,  $n_s  \approx\,
1\,-2(3\varepsilon-\eta)$,
 or equivalently
\begin{equation}
n_s  \approx\,
1\,-\frac{m_p^2}{8\pi}\,(A+V^2)^{-1/2}\,\left[1+\frac{(A+V^2)^{1/2}}{2\lambda}\right]^{-1}\,
\left(\frac{3VV'\,^2}{(A+V^2)}\,\frac{\left[1+\frac{(A+V^2)^{1/2}}{\lambda}\right]}
{\left[1+\frac{(A+V^2)^{1/2}}{2\lambda}\right]}-4V''\right).\label{nsa}
\end{equation}

One again, note that in the limit $A\rightarrow 0$, the scalar
spectral index $n_s$  coincides with that corresponding to
brane-world \cite{4}.

One of the interesting features of the five-year data set from
Wilkinson Microwave Anisotropy Probe (WMAP) is that it hints at a
significant running in the scalar spectral index $dn_s/d\ln
k=\alpha_s$ \cite{astro}. From Eq.(\ref{nsa}) we get  that the
running of the scalar spectral index becomes

\begin{equation}
\alpha_s=\left(\frac{4\,(A+V^2)}{V\;V'}\right)\,\left[\frac{\left(1+\frac{(A+V^2)^{1/2}}{2\lambda}\right)}
{\left(1+\frac{(A+V^2)^{1/2}}{\lambda}\right)}\right]\;[3\,\varepsilon_{,\,\phi}-\eta_{,\,\phi}]
\;\varepsilon.\label{dnsdk}
\end{equation}
In models with only scalar fluctuations the marginalized value for
the derivative of the spectral index is approximately $-0.03$ from
WMAP-five year data only \cite{5}.

On the other hand, the generation of tensor perturbations during
inflation would produce  gravitational waves and this
perturbations in cosmology are more involved since gravitons
propagate in the bulk. The amplitude of tensor perturbations was
evaluated in Ref.\cite{t}
\begin{equation}
{\cal{P}}_g=24\kappa\,\left(\frac{H}{2\pi}\right)^2\;F^2(x)
\simeq\frac{6}{\pi^2}\,\kappa^2\,(A+V^2)^{1/2}\,\left[1+\frac{(A+V^2)^{1/2}}{2\lambda}\right]\,F^2(x),\label{ag}
\end{equation}
where $x=Hm_p\sqrt{3/(4\pi\lambda)}$ and
$$
F(x)=\left[\sqrt{1+x^2}-x^2\sinh^{-1}(1/x)\right]^{-1/2}.
$$
Here the function $F(x)$ appeared from the normalization of a
zero-mode. The spectral index $n_g$ is given by $
n_g=\frac{d{\cal{P}}_g}{d\,\ln
k}=-\frac{2x_{,\,\phi}}{N_{,\,\phi}\,x}\frac{F^2}{\sqrt{1+x^2}}$.

From expressions (\ref{dp}) and (\ref{ag}) we write  the
tensor-scalar ratio as
\begin{equation}
r(k)=\left.\left(\frac{{\cal{P}}_g}{P_{\cal
R}}\right)\right|_{k=k_*} \simeq
\left.\left(\frac{8}{3\,\kappa}\,\frac{V'^2\;F^2(V)}{(A+V^2)\;[1+(A+V^2)^{1/2}/2\lambda]}\right)\right|_{\,k=k_*}.
\label{Rk}\end{equation} Here, $k_*$  is referred to $k=Ha$, the
value when the universe scale  crosses the Hubble horizon  during
inflation.

Combining  WMAP five-year data\cite{astro} with the Sloan Digital
Sky Survey  (SDSS) large scale structure surveys \cite{Teg}, it is
found an upper bound for $r$ given by $r(k_*\simeq$ 0.002
Mpc$^{-1}$)$ <0.28\, (95\% CL)$, where $k_*\simeq$0.002 Mpc$^{-1}$
corresponds to $l=\tau_0 k\simeq 30$,  with the distance to the
decoupling surface $\tau_0$= 14,400 Mpc. The SDSS  measures galaxy
distributions at red-shifts $a\sim 0.1$ and probes $k$ in the
range 0.016 $h$ Mpc$^{-1}$$<k<$0.011 $h$ Mpc$^{-1}$. The recent
WMAP five-year results give the values for the scalar curvature
spectrum $P_{\cal R}(k_*)\simeq 2.4\times\,10^{-9}$ and the
scalar-tensor ratio $r(k_*)=0.055$. We will make use of these
values  to set constrains on the parameters of the  model.

\section{Chaotic potential in the high-energy limit. \label{exemple}}
Let us consider  an inflaton scalar field $\phi$  with a chaotic
potential. We write for the chaotic potential  $V=m^2\phi^2/2$,
where $m$ is the mass of the scalar field. An estimation of this
parameter is given for Chaplygin standard inflation  in
Ref.\cite{Ic}. In the following, we develop models in the
high-energy limit, i.e. $\sqrt{A+V^2}\gg\lambda$.

From Eq.(\ref{NV}) the number of e-folds results in
\begin{equation}
N=\frac{2\pi}{\lambda\,m^2\,m_p^2}\;[h(V_*)-h(V_f)],
\end{equation}
where
\begin{equation}
h(V)=\frac{V^2}{2}+A\;\ln V.
\end{equation}

On the other hand, we may establish that the end of  inflation is
governed by the condition  $\varepsilon=1$, from which we get that
the square of the scalar potential becomes
\begin{equation}
V(\phi=\phi_f)=V_f=\frac{1}{2\sqrt{2\pi}}\;\left[m\,m_p\,\sqrt{\lambda}+\sqrt{\lambda\,m^2\,m_p^2-8\pi\,A}\right].
\end{equation}

Note that in the limit $A\rightarrow 0$ we obtain $V_f\sim
m\,\sqrt{\lambda}\,m_p$, which coincides with that reported in
Ref.\cite{4}.

From Eq.(\ref{dp}) we obtain that the scalar power spectrum is
given by
\begin{equation}
P_{\cal R}(k)\approx
\;\frac{8\pi}{3\,m_p^6}\,\frac{1}{m^2\,\lambda^3}\left[\frac{(A+V^2)^{3}}{V}\right],\label{ppp}
\end{equation}
and from Eq.(\ref{Rk}) the tensor-scalar ratio becomes
\begin{equation}
r(k)\approx\;\frac{4\,m_p^2\,
\lambda\,m^2}{\pi}\left[\frac{V}{(A+V^2)^{3/2}}\,F^2(V)\right].\label{rrrr}
\end{equation}

By using, that $V'\,^2=2\,m^2\,V$, we obtain from Eq.(\ref{nsa})
\begin{equation}
n_s-1=-\frac{m_p^2}{2\pi}\,\frac{\lambda\,m^2}{(A+V^2)}\left[\frac{3\,V^2}{(A+V^2)}-2\right],\label{nV}
\end{equation}
and from  Eq.(\ref{dnsdk}) that
\begin{equation}
\alpha_s\simeq-\frac{m_p^4\,m^4\,\lambda^2}{2\pi^2}\,\left[\frac{5V^2-7A}{(A+V^2)^4}\right]\,V^2.\label{as}
\end{equation}

The Eqs.(\ref{ppp}) and (\ref{nV})  has roots that can be solved
analytically for the parameters $m$ and $A$, as a function of
$n_s$, $P_{\cal R}$, $V$ and $\lambda$. The real root solution for
$m^2$, and $A$ becomes
\begin{equation}
m^2=\left(\frac{\pi}{4\,P_{\cal
R}\,\lambda^3\,m_p^6}\right)\,\left[\frac{3\,V}{2}\left(3V^4+\Im\right)+\sqrt{\Im}\left[6V^3+(n_s-1)P_{\cal
R}\lambda^2\,m_p^4\right]\right],\label{m}
\end{equation}
and
\begin{equation}
A=\frac{1}{4}\left(\sqrt{\Im}-V^2\right)\label{A}
\end{equation}
 where
 $$
\Im=9V^4+6(n_s-1)P_{\cal R}\,V\,\lambda^2\,m_p^4
 $$


From Eq.(\ref{A}) and  since $A>0$, the ratio $V^3/\lambda^2$
satisfies the inequality $V^3/\lambda^2>3(1-ns)P_{\cal
R}\,m_p^4/4$. This inequality allows us to obtain an lower limit
for the ratio $V^3(\phi)/\lambda^2$ evaluate when the cosmological
scales exit the horizon, i.e. $V_*^3/\lambda^2> 7.2\times
10^{-11}m_p^4$. Here, we have  used the WMAP five year data where
$P_{\cal R}(k_*)\simeq 2.4\times 10^{-9}$ and $n_s(k_*)\simeq
0.96$.

 Note that
in the limit $A\rightarrow 0$,  the constrains
$m\approx10^{-5}M_5$ and $V_*=m^2\phi_*^2/2\approx10^{-4}M_5^4$
are recovered \cite{4}. Here,  we used the relation
$m_p=M_5^3\,\sqrt{3/(4\pi\lambda)}$.

 In Fig. 1 we
have plotted the adimensional quantity   $\lambda^2/m_p^8$ versus
the adimensional scalar potential evaluated when the cosmological
scales exist the horizon $V_*/m_p^4$. In doing this, we used
Eq.(\ref{as}) that has roos that can be solved for the brane
tension $\lambda$, as a function of $\alpha_s$, $m$, $A$ and $V$.
For a real root solution for $\lambda$, and from Eqs. (\ref{m})
and (\ref{A}) we obtain a relation of the form $\lambda=f(V_*)$
for a fixed  values of $\alpha_s$, $n_s$ and $P_{\cal R}$. In this
plot we using the WMAP five year data where $P_{\cal R}(k_*)\simeq
2.4\times 10^{-9}$,  $n_s(k_*)\simeq 0.96$,
$\alpha_s(k_*)\simeq-0.03$. In Fig. 2 we have plotted the
tensor-scalar ratio given by Eq.(\ref{rrrr}) versus the
adimentional parameter $A/m_p^4$.  The WMAP five-year data favors
the tensor-scalar ratio $r\simeq 0.055$  and the from Fig. 2 we
obtain that $A$ parameter becomes $A\simeq 4\times10^{-13}m_p^8$.
For this value for $A$ parameter we  get the values
$V_*=4\times10^{-7}m_p^4$,  $\lambda\simeq 2\times 10^{-8}m_p^4$,
and $m\simeq  10^{-5}m_p$. Also,  the number of e-folds, $N$,
becomes of the order of $N\sim 55$.  We should note also that the
$A$ parameter becomes  large     by two order of magnitude and the
$m$ parameter becomes similar  when it is compared with the case
of Chaplygin inflation in the low-energy limit\cite{Ic}.

 On the other hand, is interesting to compare the role that brane
effects play in our model, with the  one they play in the context
of tachyonic inflation \cite{x}. In doing this, we using the above
parameters, i.e. $\lambda\simeq 2\times 10^{-8}m_p^4$ and
$V_*=4\times10^{-7}m_p^4$ which correspond to the five-dimensional
Planck mass, $M_5\simeq6.6\times10^{-2} m_p$ and $\phi_*\simeq
89m_p$. Here, we used the relations
$m_p=M_5^3\,\sqrt{3/(4\pi\lambda)}$, $V_*=m^2\phi_*^2/2$ and
$m\sim10^{-5}m_p$.
 We
should note that $M_5$ becomes similar and the $\phi_*$ scalar
field becomes  large  by two order of magnitude when it is
compared with the case of tachyonic inflation.  In addition, the
number of e-folds, $N$, is smaller than the reported in tachyon
inflation.

\begin{figure}[th]
\includegraphics[width=3.0in,angle=0,clip=true]{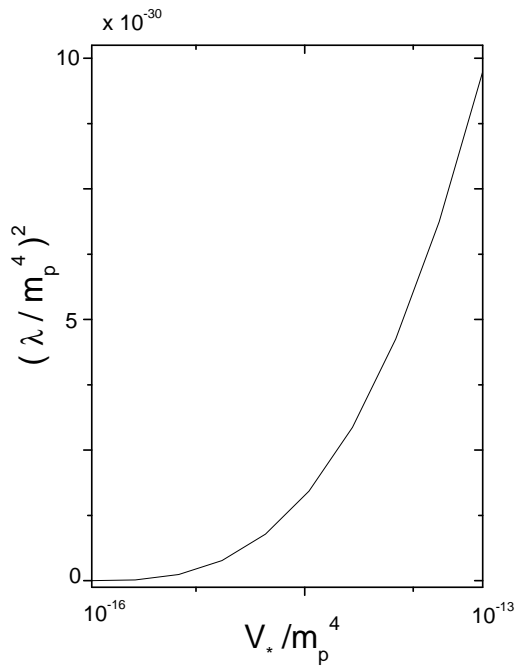}
\caption{The plot shows the adimentional square of the brane
tension $(\lambda/m_p^4)^2$ versus the adimentional scalar
potential
 $V_*/m_p^4$. Here, we have used the WMAP five-year data where
$P_{\cal R}(k_*)\simeq 2.4\times 10^{-9}$, $n_s(k_*)\simeq 0.96$
and $\alpha_s(k_*)\simeq-0.03$.
 \label{rons}}
\end{figure}

\begin{figure}[th]
\includegraphics[width=3.0in,angle=0,clip=true]{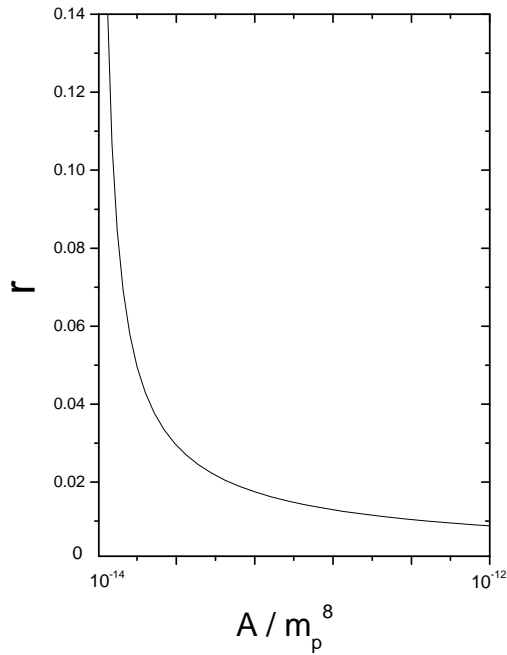}
\caption{The plot shows the tensor-scalar ratio $r$ versus the
adimentional parameter
 $A/m_p^8$. Here, we have used  the WMAP five-year data where
$P_{\cal R}(k_*)\simeq 2.4\times 10^{-9}$, $n_s(k_*)\simeq 0.96$
and $\alpha_s(k_*)\simeq-0.03$.
 \label{rons}}
\end{figure}


\section{Conclusions \label{conclu}}

In this paper we have studied the brane-Chaplygin inflationary
model. In the slow-roll approximation we have found a general
relation between the scalar potential and its derivative. This has
led us to a general criterium for inflation to occur (see
Eq.(\ref{cond})). We  have also obtained   explicit expressions
for the corresponding scalar spectrum index $n_s$ and its running
$\alpha_s$.

 By using  a chaotic potential in the high-energy
regime  and from the WMAP five year data,  we found the
constraints  of the parameter $A$ from the tensor-scalar ratio $r$
(see Fig. 2).  In order to bring some explicit results we have
taken the constraints $A\simeq 10^{-13} m_p^8$, from which we get
the values  $V_*\simeq 3\times 10^{-7}m_p^4$, $\lambda\simeq
2\times 10^{-8}m_p^4$  and $m\simeq 10^{-5}m_p$. Here, we have
used  the WMAP five year data where $P_{\cal R}(k_*)\simeq
2.4\times 10^{-9}$, $n_s(k_*)\simeq 0.96$,
$\alpha_s(k_*)\simeq-0.03$ and $r(k_*)\simeq 0.055$. Note that the
restrictions imposed by currents observational data allowed us to
establish a small range for the parameters that appear in the
brane-Chaplygin inflationary model.

We have not addressed reheating and transition to standard
cosmology in our model (see e.g., Ref.\cite{u}). However, a more
accurate calculation for the reheating temperature  in the
hight-energy scenario, would be necessary for establishing some
constrains  on the parameters of the model. We hope to return to
this point in the near future.

\begin{acknowledgments}
  This work  was supported by the ``Programa Bicentenario de
Ciencia y Tecnolog\'{\i}a" through the Grant \mbox {N$^0$ PSD/06}.
\end{acknowledgments}


\end{document}